\documentclass{emulateapj}

\usepackage{graphicx,xspace,floatflt}
\usepackage{amssymb,amsmath}
\usepackage[normalem]{ulem}
\usepackage{color}
\begin{document}

\title{The Impact of Accurate Extinction Measurements for X-ray Spectral Models}
\author{Randall K. Smith\altaffilmark{1}, Lynne A. Valencic\altaffilmark{2,3}, Lia Corrales\altaffilmark{4} }
\altaffiltext{1}{Smithsonian Astrophysical Observatory, 60 Garden Street, Cambridge, MA 02138, USA}
\altaffiltext{2}{NASA Goddard Space Flight Center, Greenbelt, MD 20771, USA; lynne.a.valencic@nasa.gov}
\altaffiltext{3}{The Johns Hopkins University, Department of Physics \& Astronomy, 366 Bloomberg Center, 3400 N. Charles St., Baltimore, MD 21218, USA}
\altaffiltext{4}{MIT Kavli Institute for Astrophysics and Space Research, 77 Massachusetts Ave, 37-241, Cambridge, MA 02139, USA}

\begin{abstract}
Interstellar extinction includes both absorption and scattering of photons from interstellar gas and dust grains, and it has the effect of altering a source's spectrum and its total observed intensity. However, while multiple absorption models exist, there are no useful scattering models in standard X-ray spectrum fitting tools, such as XSPEC. Nonetheless, X-ray halos, created by scattering from dust grains, are detected around even moderately absorbed sources and the impact on an observed source spectrum can be significant, if modest, compared to direct absorption. By convolving the scattering cross section with dust models, we have created a spectral model as a function of energy, type of dust, and extraction region that can be used with models of direct absorption. This will ensure the extinction model is consistent and enable direct connections to be made between a source's X-ray spectral fits and its UV/optical extinction.
\end{abstract}

\maketitle

\section{Introduction}

Determining the true spectrum of an astronomical source requires correction from the effects of passing through the interstellar medium (ISM), typically termed ``extinction.'' Both absorption and scattering, each of which are energy-dependent, sum to create the total extinction.

Depending upon their energy, photons might be absorbed primarily by atoms, molecules, or dust.  However, for frequencies above radio, scattering occurs primarily due to interactions with dust (although resonant line scattering can also occur).  In the UV/optical regime, scattering from dust grains changes the photon direction dramatically, entirely removing it from the beam.  At X-ray energies small-angle scattering dominates, creating an arcminute-scale X-ray ``halo'' around bright sources with significant dust along the line of sight.  Just as with optical scattering, this effect can and does impact the source spectrum.

Predicted by \citet{Overbeck65}, the first X-ray halo was detected with the Einstein IPC around the source GX339-4 \citep{Rolf83}.  \citet{PS95} used ROSAT to find halos around 28 sources. \citet{VS15} (hereafter VS15) updated this survey using Chandra and XMM-Newton data and found X-ray halos around a number of moderately absorbed (N$_{\rm H} \sim 3\times10^{21}$\,cm$^{-2}$)\ point sources. 

In general the interaction of any photon with a spherical grain can be treated using the exact Mie solution.  However, in some cases X-ray scattering in dust can be approximated as Rayleigh scattering, leading to a simple analytic solution. For sufficiently small grains, photon wavelengths, and scattering angles the scattering is coherent ($\propto n_e^2$), so small-angle scattering dominates the total scattering. By integrating the scattering cross section over the entire grain we obtain the Rayleigh-Gans (RG) approximation \citep{vdH57}.   An exact analytic solution exists for spherical dust particles \citep{ML91}.  If the optical constants are taken from the Drude approximation \citep{BH83}, the core behavior of the solution, at small angles where the scattering cross section is maximal, can be fit with a Gaussian function that provides insight into the relevant scales involved:
\begin{equation}
{{d\sigma}\over{d\Omega}}(E,~a,~\theta_{\rm sca})~\propto~\rho^2 a^6
\exp(-0.4575 E^2 a^2 \theta_{\rm sca}^2)
\label{RGdiffsigma}
\end{equation}
where $E$\ is the X-ray energy in keV, $a$\ the dust radius in $\mu m$, $\rho$\ the dust grain density in g cm$^{-3}$, and $\theta_{\rm sca}$\ the scattering angle in arcminutes. This equation (along with a similar approximation to the size of the halo) is the basis of the XSPEC \citep{Arnaud96} X-ray scattering model (named ``dust'').  However, as shown by \citet{SD98}, the RG approximation fails right where the halo is strongest, when modeling low-energy ($< 1$\,keV) X-rays scattering from realistic interstellar dust distributions.  Complicating matters further, the XSPEC ``dust'' model normalization is arbitrary -- it does not specify any particular dust model -- and so cannot be compared to the direct absorption or to an optical/UV extinction measurement.  

Regardless of these difficulties, dust scattering impacts the observed source spectrum significantly, with details depending upon the energy, type of dust, and extraction region employed. Although the first exact calculations of the scattering cross section as a function of energy from astrophysical dust were done by \citet{Draine03} over a decade ago, observers have to date considered the impact of dust scattering on spectra via a phenomenological approach. \citet{Ueda10} used a model based on scaling the energy-dependent halo profile observed around GX13+1 \citep{SES02}, which was itself fit using an earlier version of the \textit{xscat} code.  We have now extended the \textit{xscat} code to perform full Mie calculations of the scattering cross section combined with published interstellar dust models to create a realistic model for the dust scattering that can be used with existing models of direct absorption to determine the true extinction along the line of sight.  

\section{Method}

\begin{figure}
\includegraphics[totalheight=2.in]{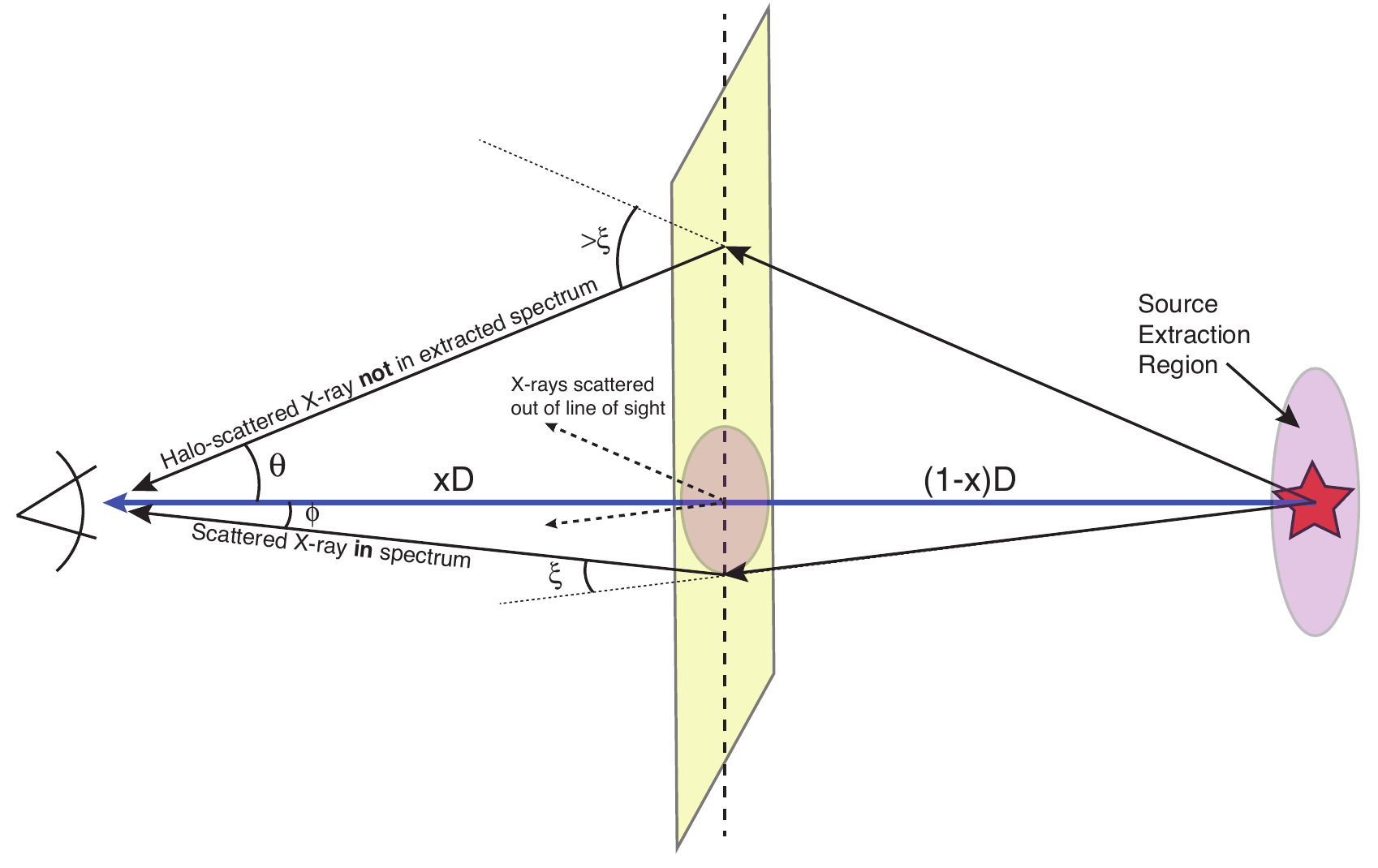}
\caption{The geometry of X-ray scattering from a plane of dust between the source and the observer. The X-ray scattered at the top of the figure will appear in X-ray halo around the source, but will not be included in the source's extracted spectrum.  However, for a given extraction region of radius $\phi$, there will exist a minimum scattering angle $\xi$ ($\approx \phi/(1-x)$) for small values of $\phi/(1-x)$ such that any X-ray scattered through a larger angle will be excluded from the spectrum.\label{fig:ScatDiag}}
\end{figure}

The basic geometry of the scattering for the case of plane of dust between the source and the observer, together with the impact of a finite extraction region is shown in Figure~\ref{fig:ScatDiag}.  The distance to the dust cloud is $xD$, where $D$ is the total distance to the source.  As the angles involved in X-ray scattering are small, simplifying approximations can be made \citep{SD98} between the observed angle and the actual scattering angle.  

The total cross section $\sigma_{LOS}(E, \phi)$\ for an X-ray of energy $E$\ scattering out of an extraction region of size $\phi$ can be written as
\begin{equation}
\begin{split}
\sigma_{LOS}(E, \phi)  & =  {\rm N}_{\rm H} \int n(a) da \times \\ 
  & \int_{\xi}^{\xi^{\prime}} 2 \pi \sin(\theta)  {{d\sigma}\over{d\Omega}}(E, \theta, a, x)  d\theta
 \end{split}
 \label{eq:sigma}
\end{equation}

where $n(a)$\ is the dust size distribution for grains of size $a$, normalized by the hydrogen column density N$_H$, and $d\sigma/d\Omega$\ is the X-ray scattering cross section itself.  The formal limits of integration are $\xi = \arctan((x \tan\phi)/(1-x))$\ and $\xi^{\prime} = \pi$, but as the scattering angles are small, this can be simplified to $\xi = {\phi}/{(1-x)}$ and $\xi^{\prime} \sim 1^{\circ}$; further simplification can be made in the integrand with the small-angle approximation $\sin(\theta) \sim \theta$.

We focus on scattering from a single dust plane -- most likely a molecular cloud along the line of sight -- as the VS15 survey of X-ray halos found that this is the most common solution when fitting X-ray halo profiles. The narrow annuli seen in rings of scattered X-rays created by absorbed variable sources \citep[\protect{{\it e.g.}}][]{Vaughan04, Tiengo10} provides additional evidence, as a smooth dust distribution lit by a suddenly-brighter source would generate a filled circle of whose radius increased with time.

If desired, however, a set of $\sim 5$\ clouds spaced evenly along the line of sight provides a close approximation to a smooth distribution.  This was determined by examining those halos in VS15  that could be fit with a smooth distribution. The data were analyzed following the method of VS15, and interested readers are referred to that work for more information.

\begin{figure}
\includegraphics[totalheight=2.0in]{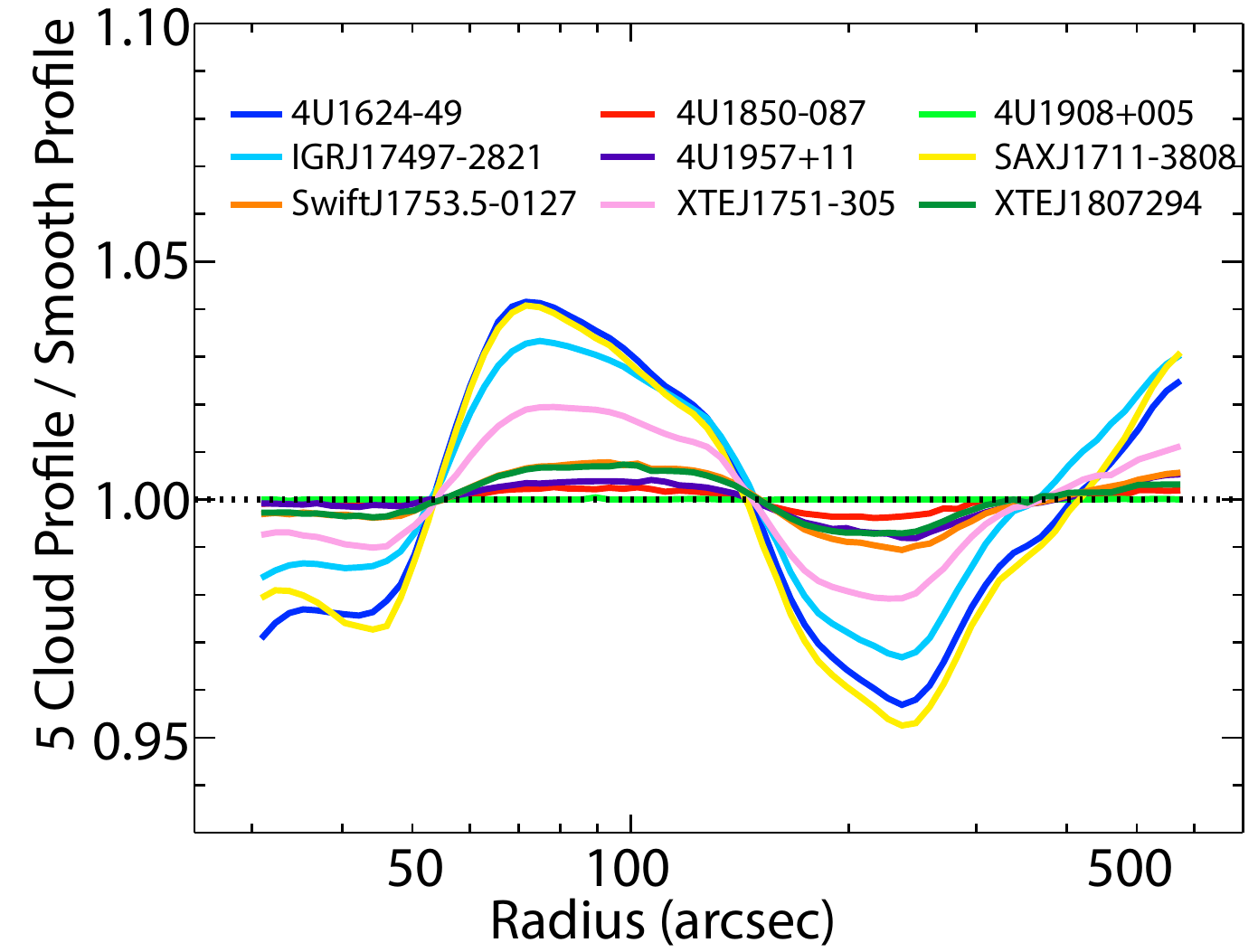}
\caption{The ratio of the halo radial surface brightness fits for a 5 cloud distribution compared to the same profile for a smooth distribution.  The variations are typically less than 5\%. \label{fig:5cl}}
\end{figure}

The radial profiles were fit with the \citet{MRN77} model twice. First, they were fit assuming a smooth distribution, allowing N$_{\rm{H}}$ to float. Then, they were fit using 5 evenly-spaced clouds, holding the N$_{\rm{H}}$\ of each cloud at 1/5 the value found for the smooth distribution. The ratio of the fits for a subset of these sources, chosen to cover a wide range of N$_{\rm{H}}$, is shown in Figure \ref{fig:5cl}. The similarity between model fits follows N$_{\rm{H}}$ closely, with $< 1\%$
difference for lightly absorbed sources (N$_{\rm{H}} < 3\times10^{21}$\,cm$^{-2}$), and the highest deviations (about 3-5\%) seen toward the most heavily absorbed sight lines (N$_{\rm{H}} \sim 10^{22}$cm$^{-2}$. The values of $\chi^2$ for these fits are listed in Table \ref{tab:chi2list}, as is the HI column density within a 1$^{\circ}$ radius of the source \citep{LAB05}.

\begin{table}
\caption{Comparison of $\chi^2$ and N$_{\rm{H}}$\label{tab:chi2list}}
\begin{tabular}{lccc}
\tableline\tableline
Source  & $\chi^2$(smooth) & $\chi^2$(5cloud) & N$_{\rm{H}}^{a}$  \\
\tableline
4U 1957+11         & 5.92 & 5.62 & 1.2 \\
Swift J1753.5-0127 & 1.25 & 1.21 & 1.7 \\
4U 1850-087        & 1.43 & 1.40 & 2.4 \\
XTE J1807          & 1.32 & 1.28 & 2.5 \\
4U 1908+005        & 1.25 & 1.23 & 2.8 \\ 
XTE J1751-305      & 0.88 & 0.90 & 6.3 \\
SAX J1711-3808     & 2.69 & 3.29 & 11 \\ 
IGR J17497-2821    & 3.28 & 2.49 & 12 \\
4U 1624-49         & 11.3 & 16.0 & 16 \\
\tableline
\tableline
\end{tabular}

$^a${In units of $10^{21}$\ cm$^{-2}$. From Kalberla et al. (2005).}
\end{table}

\subsection{Scattering Cross Sections}

%The basic scattering cross section and halo size for X-rays interacting with grains of any size or composition (including grains with icy mantles). The code is called `xscat' and is available to the community\footnote{http://hea-www.harvard.edu/$\sim$rsmith/software/xscat-1.0.0.tar.gz}.
Considering scattering through dust simply as a wave interacting with a sphere, and given the difficulties with the RG approximation, the only effective approach is to use the exact Mie solution with a specialized code that can handle the large size parameters ($x \equiv 2\pi a/\lambda \sim 10,000$) involved. We use the Mie code developed by \citet{Wiscombe79, Wiscombe80}, which was written for atmospheric use but has been tested for size parameters up to 20,000, sufficient for our problem\footnote{This code is available at ftp://climate1.gsfc.nasa.gov, in the directory wiscombe/Single\_Scatt/Homogen\_Sphere/Exact\_Mie/}.

A Mie code requires the use of accurate optical constants $(m = n+ik)$. For the dust models described below we used the precalculated optical constants from \citet{ZDA04}, who used this same procedure to extend a range of typical dust components well into the hard X-rays.  These values are also provided in the {\it xscat}\ software package. The \citet{ZDA04} optical constants are ultimately based on the photoionization cross section compilation by \citet{Verner96} used in conjunction with the Kramers-Kroenig relation to derive a consistent value for $n$. 

These optical constants do not include detailed X-ray Absorption Fine Structure (XAFS) effects  \citep[\protect{\it e.g.}][]{Lee09} that are present around atomic edges.  The intent of {\it xscat}\ is to provide a robust measure of scattering over a broad bandpass.  The simple edges in \citet{Verner96} are adequate to diagnose basic dust parameters such as relative abundances or compositions \citep{HD15}.  Using an observatory with sufficient angular and spectral resolution to resolve scattered photons around an edge with XAFS, however, would allow detailed studies of grain mineralogy and geometry \citep{HD15}.  For example, the {\sl Athena} X-ray Integral Field Unit \citep{Nandra13} will provide $5''$\ angular and 2.5 eV spectral resolution, easily enough to detect the effect of both absorbed and scattered photons near an edge with XAFS.

\subsection{Dust Size Distributions}

\begin{figure}
\includegraphics[totalheight=2.0in]{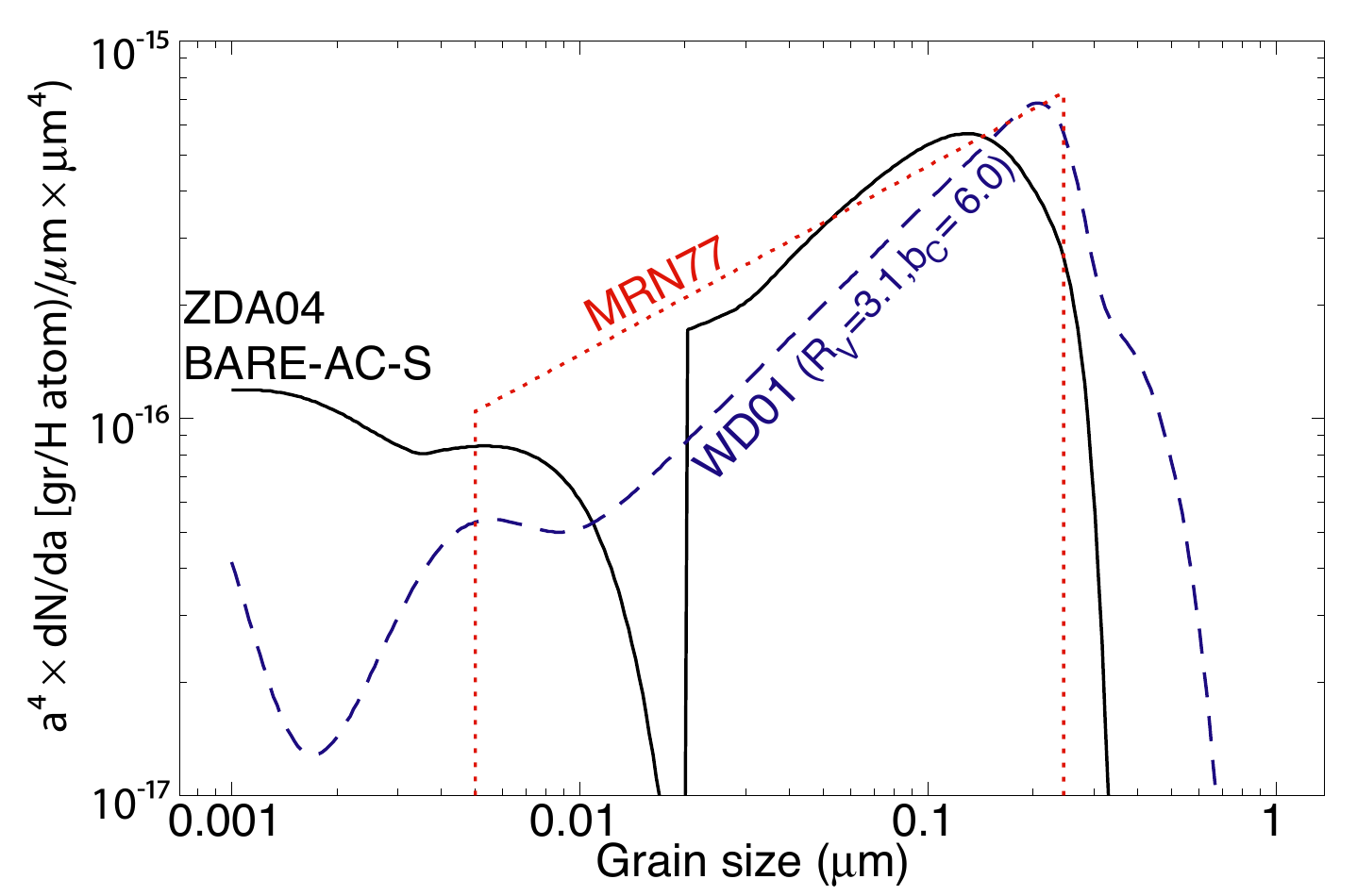}
\caption{The total dust size distribution of all model components, weighted by the dust radius to the fourth power to show which grain volumes (i.e. masses) per unit (log) radius dominate the distribution.  While similar, these dust models predict quite different scattering as a function of energy and extraction region. 
\label{WtSizeDist}}
\end{figure}

Determining the impact of dust on X-ray spectra requires more than calculating the scattering cross section integrated over a range of angles for grains of a single size and composition.  As Eq.~\ref{eq:sigma} shows, the cross section must be combined with a dust model that specifies the grain size distribution and composition.  A large number of such models exist; at UV/optical wavelengths these include Mathis, Rumpl \& Nordsieck (1977; MRN77), Weingartner \& Draine (2001; WD01), and Zubko et al. (2004; ZDA04).  The latter two papers actually include a wide range of models, so these three papers themselves include almost 100 different models (and over 2500 citations). Figure~\ref{WtSizeDist} shows the differences in grain size distribution for just three of these models, weighted by the dust radius to the fourth power -- proportional to the total scattering cross section in the RG approximation.  The differences between these models can be extremely significant to the final results.

\subsection{Calculations\label{sec:calc}}

The \textit{xscat} code calculates values of $\sigma_{LOS}(E,\phi)$\ for a range of dust models, performing a separate calculation for each grain component in the model ({\it e.g.} silicate, graphite, composite, PAH, etc).  We considered a range of relative dust cloud positions including $x=0$, 0.2, 0.5, 0.75, 0.9, 0.95, 0.99, and 0.999 where the observer is at x=0 and the source is at $x=1$.  For a fixed source extraction radius, as the dust cloud gets close to the X-ray source, the excluded scattering angle range increases.  This creates a vanishingly small halo that becomes increasingly slow to integrate numerically. For a source at 10 Mpc, a value of x=0.999 corresponds to a cloud 1 kpc distant from the source, allowing model to include dust in nearby galaxies.  We also ran the models from energies between 0.1-3 keV, in steps of 2 eV; at energies higher than 3 keV, the RG approximation is adequate.  Complete runs of three dust models were generated, including MRN77, WD01 (Galactic dust, case A, $R_V=3.1$, $b_C=6.0$), and the ZDA04 model with bare grains, amorphous carbon, and solar abundances (ZDABAS). Other models can be calculated as desired; typical runtimes are $\sim 1$\,week on a modern computer.

We also compared the results of \textit{xscat} against the model shown in Figure 6 of \citet{Draine03}, with agreement at the 10-50\% between 250-800 eV and $<10$\% above 800 eV.  Tests show that the differences at low energies are due to the different optical constants used. It should be noted that \citet{Draine03} includes both grain scattering and absorption with detailed edge effects as well as gas-phase absorption; {\it xscat}\ only calculates grain scattering, and as noted above uses optical constants with simple edges.  This is to maintain consistency with existing XSPEC absorption models {\it  e.g. phabs, tbabs}\ that use these optical constants and in the case of {\it tbabs} already include grain absorption \citep{Wilms00}. Although beyond the scope of this paper, interpreting high-resolution X-ray spectra will require a self-consistent extinction model that  includes a plausible range of interstellar dust models and self-consistent optical constants. We plan to complete this work in a subsequent paper.  

\begin{figure*}
\includegraphics[totalheight=2.2in]{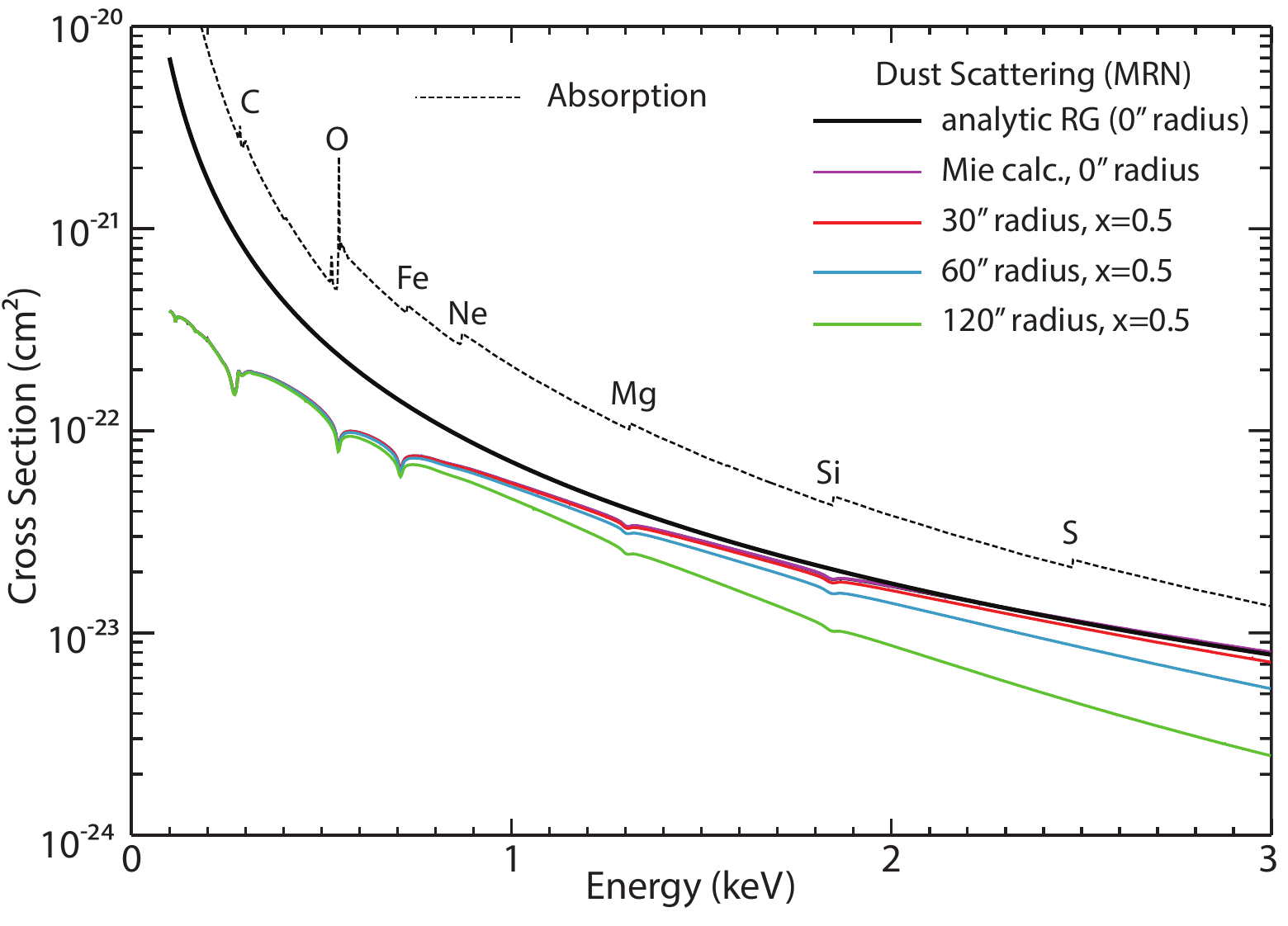}
\includegraphics[totalheight=2.3in]{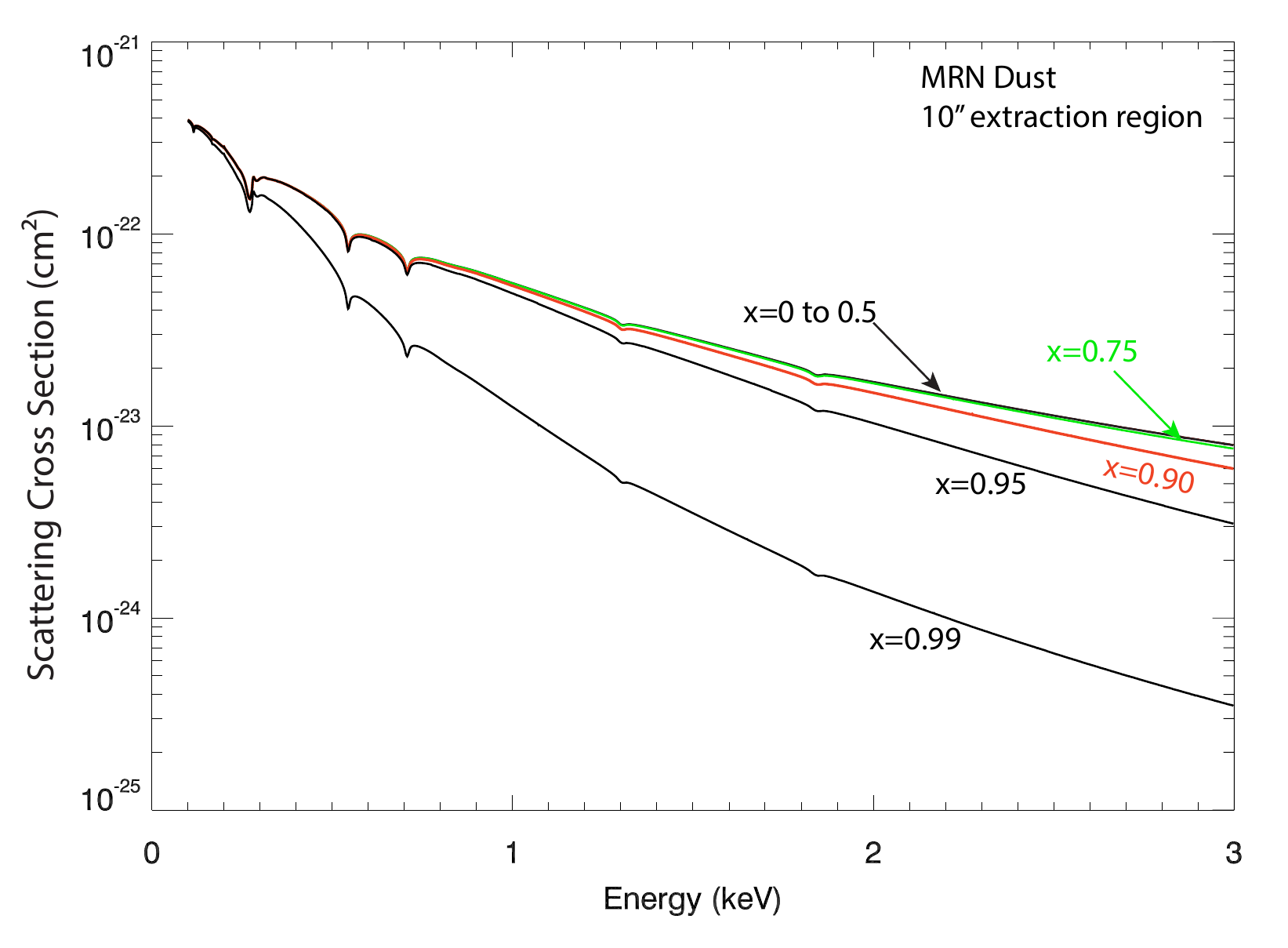}
\caption{[Left] A comparison of the ISM X-ray scattering and absorption cross sections. The dotted curve shows the \protect{\citet{Verner96}} photoionization cross sections.  A range of MRN-type dust scattering cross sections are shown for a dust cloud midway between source and observer ($x=0.5$) and $0'', 30'', 60'',$ and $120''$\ extraction radii.  Also shown for comparison is the analytic RG approximation with $0''$\ radius extraction region.  Although scattering is not the dominant term, it is also not negligible.  [Right] The scattering cross section as a function of dust position along the line of sight, all for a $10''$\ extraction region, showing that position only matters for dust very near the source. \label{fig:sigma}}
\end{figure*}

The output from each collection of runs was combined into a single FITS file, which can be read by the newly developed \textit{xscat} XSPEC model (also provided as part of the \textit{xscat} package).  Based on the user-input parameters this code reads the appropriate file, and determines the scattering cross section either by interpolating between the calculated energies or extrapolating using an RG model. Figure~\ref{fig:sigma} shows some sample results from these runs. 

As Figure~\ref{fig:sigma} shows, the spectrum of the scattered X-rays exhibits features resulting from the K-edge absorption of oxygen (0.532 keV) and the L-edge of iron ($\sim$\,0.7 keV), due to silicate components in the dust; the strong features from oxygen are observable at CCD-resolution ($\Delta E \approx 100$\,eV).  Figure~\ref{fig:sigma}[Left] also demonstrates that while the RG approximation is useful at high energies, it is inadequate at energies where the scattering is significant. 

%Dust Models:  MRN, WD3100AGAL, ZDABAS %%%, ZDABAF, ZDABAB, ZDABGS, ZDACGS

\section{Effects of Dust Scattering on Selected Sources}

We examined the impact of dust scattering on a range of sources, using an XSPEC model (also named \textit{xscat}) that uses the output of the \textit{xscat} code described in \S\ref{sec:calc}.  This XSPEC model is available at \url{https://github.com/AtomDB/xscat}.  Typical  results of including dust scattering are shown below.

\subsection{Cooling Neutron Stars}

\subsubsection{XTE J1701-462}

XTEJ1701-462 is a neutron star binary system at Galactic $(l,b) = (340.81, -2.488)$\ and an estimated distance of 8.8 kpc \citep[hereafter F10]{Fridriksson10}.   At this distance and position, the source is $\sim 380$\,pc out of the plane, suggesting most of the absorption is in the foreground.  On 2006 January 18 the source went into a super-Eddington outburst state \citet{RemillardLin06}, which lasted for $\sim 1.5$\,years.  F10 fit a series of observations made after the end of the outburst in order to measure the temperature and flux from the neutron star as it cooled. They used a spectral model consisting of an absorbed NS atmosphere model plus a powerlaw (an XSPEC model of {\it phabs*(nsatmos + pegpwrlw)}).  The power-law term was not physically motivated, but rather based on extensive experience with non-thermal components in X-ray binary spectra.  A joint fit to the complete data set was used to determine reasonable values for the constant parameters, including distance (8.8 kpc), radius (10 km), mass (1.4 M$_{\sun}$), absorbing column ($1.93\pm0.02\times10^{22}$\,cm$^{-2}$), and power-law slope ($1.93\pm0.2$).  

We reanalyzed Chandra ObsID 7515, F10's 'CXO-3' that was observed $\sim 174$\ days after the end of the outburst, using CIAO 4.7 (CALDB 4.6.8) to see the impact of including dust scattering.  For the non-time-varying parameters in the fit we used the values as given above, and fit only for the temperature and flux of the neutron star and the flux in the power-law component.  In Table~\ref{tab:XTEJ1701} we show the values from F10 (who used CIAO 4.2 and CALDB 4.2.0), along with our values using the same model, and the values after including a dust scattering term. The dust scattering model assumed MRN77 dust, a cloud distance of 4.4\,kpc (corresponding to a height above the Galactic disk of $\sim 190$\,pc), and an equivalent hydrogen column density of $8\times10^{21}$\,cm$^{-2}$.  This latter value is roughly midway between the  $7.19\times10^{21}$\,cm$^{-2}$\ from the \citet{LAB05} survey, or $8.8\times10^{21}$\,cm$^{-2}$\ from the \citet{DL90} survey, based on the HEASARC N$_{\rm H}$\footnote{http://heasarc.gsfc.nasa.gov/cgi-bin/Tools/w3nh/w3nh.pl} tool.  The F10 values were determined by a simultaneous fit to 13 separate spectra, tying the absorbing column density and power law slope values together for all fits, but allowing the parameters shown in Table 2 to float. As this project is only intended to show variations due to scattering, we held the absorbing column density and power law slope constant at F10's best-fit values of N$_{\rm H} \equiv 1.93\times10^{22}$\,cm$^{-2}$\ and $\Gamma \equiv 1.93$. 

When using the same model, data binning and statistical method as F10, we found similar values for the neutron star effective temperature, bolometric luminosity, and power law flux, although the temperature values had much larger errors (presented as 1$\sigma$\ values here to match F10).  The reason for this latter discrepancy is unclear.  We attempted to match the extraction regions used by F10, but insufficient details exist to be certain of an exact match.  In addition, we are using updated software and calibration files, which will affect the results  With only $\sim 500$\, counts in the source spectrum, achieving $<4$\% accuracy on the temperature would seem difficult, but might be due to some impact of the simultaneous fit of 13 datasets used in F10.  

\begin{table}
\caption{Best-fit parameters for the cooling neutron star XTEJ1701-462 at day 174 after the end of outburst\label{tab:XTEJ1701}}
\begin{tabular}{llll}
\hline \hline
Source             &  $kT^{\infty}_{\rm eff}$[eV] &  $F_{bol}^a$ & $F_{pl}^a$ \\ \hline
F10                  & $129.1\pm4.7$ & $6.6\pm1.0$  & $4.8\pm0.9$ \\
This paper       & $136\pm28$     & $8.6\pm2.2$ & $4.5\pm0.5$ \\
w/scattering     & $126\pm23$     & $12.3\pm3.3$ & $4.6\pm0.5$ \\ \hline
\end{tabular}

$^a$\,in units of $10^{-13}$\,erg\,cm$^{-2}$\,s$^{-1}$ \\
\end{table}

When including dust scattering, however, we see that the best-fit temperature drops by $\sim 0.5\sigma$, while the neutron star flux increases by 43\% from the previous best fit value.  In part, these changes are due to holding the absorbing column density fixed -- had this been allowed to vary, it might have resulted in a smaller increase in flux.  The statistics of the fit change minimally after adding the scattering model, which itself has no free parameters. Without redoing the full analysis done in F10, it is impossible to determine the ultimate impact of dust scattering on the cooling term, except to note that it could easily change the cooling parameters by 1-2$\sigma$. 

\subsubsection{EXO 0748-676}

Similar to XTEJ1701-462, EXO 0748-676 is an X-ray binary that transitioned from a long (24 yr) outburst phase to relative quiescence in 2008, inspiring  \citet{Degenaar11}[hereafter D11] to study the cooling of the crust.  Unlike XTEJ1701, however, the Galactic line of sight hydrogen column density towards the source is nearly an order of magnitude lower at $\sim 10^{21}$\,cm$^{-2}$ \citep{LAB05,DL90}.  Similar to F10, D11 considered a wide range of observations to measure the cooling, fitting the same model consisting of an absorbed neutron star atmosphere plus a power law to each (in XSPEC, {\it phabs*(nsatmos + powerlaw)}.  

We selected the first observation described in D11, an XMM-Newton observation (ObsID 0560180701) and reanalyzed the dataset using SAS v14.0 and the most recent calibration database.  The data were extracted following the methods and extraction regions described in D11, although they used SAS v9.0 and an earlier calibration database. The MOS1, MOS2, and pn data were all fit jointly, with a linear scaling applied to pn data to allow for calibration uncertainties (the value was $1.02\pm0.03$).  The best-fit parameters are shown in Table~\ref{tab:EXO0748v1}, where $F_{bol}$\ is the unabsorbed neutron star flux between 0.01-100 keV and $F_X$\ is the total 0.5-10 keV model flux.  We used the same values for the constant terms as D11, including N$_{\rm H} = 7\times10^{20}$\,cm$^{-2}$, M$_{\rm NS} = 1.4M_{\sun}$, R$_{\rm NS} = 15.6$\,km, D=7.4 kpc, and a power-law index of $\Gamma=1.7$. As with the results for XTEJ1701-462 shown in Table~\ref{tab:XTEJ1701}, the impact of including dust scattering is modest, but larger than the statistical uncertainties.  As with the F10 results, we find good agreement with the flux measurements when using the same model, but here the best-fit neutron star effective temperature is higher as well as having larger error bars (in this case, 90\% limits to match the results reported in D11. After adding in the dust scattering, assumed to be MRN77 dust positioned halfway to the source with the same total equivalent hydrogen column density, the neutron star's temperature drops and its bolometric flux increases.  The change is not enough to invalidate any results, but it is both significant and impacts the fits systematically. 

\begin{table}
\caption{Best-fit parameters for the cooling neutron star EXO 0748-676 observed on 2008-11-06\label{tab:EXO0748v1}}
\begin{tabular}{llll}
\hline \hline
Source               &  $kT^{\infty}_{\rm eff}$[eV] & $F_X^a$ &$F_{bol}^a$  \\
%			& eV     & $10^{-12}$\,cgs$^a$ & $10^{-12}$\,cgs \\
%\protect{\citet{Degenaar11}} & $120.7\pm0.4$ &$1.14\pm0.01$ & $1.39\pm0.02$   \\
D11                    & $120.7\pm0.4$ &$1.14\pm0.01$ & $1.39\pm0.02$   \\
This paper         & $128.6\pm2.6$ &$1.15\pm0.02$ & $1.41\pm0.04$\\
w/scattering       & $126.0\pm2.5$ & $1.19\pm0.03$ & $1.48\pm0.04$\\ \hline
\end{tabular}

$^a$\,in units of $10^{-12}$\,erg\,cm$^{-2}$\,s$^{-1}$ \\
\end{table}

\subsection{Black Hole Binaries: GRS1758-258}

Dust scattering does not only impact parameters measured for cooling neutron stars.  GRS1758-258 is a Galactic microquasar, a stellar mass black hole in a binary system, but the high extinction to the system \citep{Rothstein02} has made unambiguous identification of its companion star and period difficult.  \citet{Soria11} analyzed a series of three XMM-Newton observations of the system between 2000-2002, and found that in 2001 (ObsID \#0136140201) the source was in the disk-dominated `soft' state.  As soft X-rays are most affected by dust scattering, we chose to re-analyze this dataset to determine the impact of adding {\it xscat}\ to the spectral model.  

We re-analyzed the XMM-Newton data using SAS v14.0 (\citet{Soria11} used v10.0), but otherwise followed the extraction approach outlined in their paper.  We compare only to the MOS1 data, as \citet{Soria11} noted that MOS2 was affected by ``anomalously low count rate'' regions near the source, while the pn was seriously affected by pileup.  The fit used the same model, a disk blackbody plus a power-law component, including both a Galactic and a intrinsic absorption component (e..g, $ phabs \times phabs \times (diskbb + pow)$ in XSPEC).  In Table~\ref{tab:GRS1758} we show the best-fit parameters from \citet{Soria11}, our best-fit values with the same model, and our best-fit values including dust scattering.  The {\it diskbb}\ parameters include the disk inner temperature ($kT_{dbb}$) and normalization ($N_{dbb}$), along with the power-law model's slope ($\Gamma$) and normalization ($N_{po}$). In all of our fits, we found it necessary to fix the slope of the power-law to the value found in \citet{Soria11} in order to reasonably constrain the fits; this may be due to updates in the XMM-Newton calibration.  Holding the slope constant also had the effect of artificially reducing the size of the $kT_{dbb}$\ error bars, but this is not relevant to our comparison here. With the slope fixed we obtained similar values as \citet{Soria11}. 
\begin{table}
\caption{Best-fit spectral parameters for GRS1758-258 \label{tab:GRS1758}}
\begin{tabular}{llll}
\hline \hline
Parameter & Soria+(11)  & This paper & +xscat \\ \hline
$N_H$(gal, abs)$^a$ & $0.75$                           &	0.75				& 0.75 \\
$N_H$(gal, scat)$^a$ & --                                   &					& 0.75 \\
$N_H$(int)$^a$          & $0.99^{+0.04}_{-0.02}$ &$1.02\pm0.02$	& $0.77\pm0.02$		\\
$kT_{dbb}$\ (keV)&$0.45^{+0.01}_{-0.01}$  &$0.447\pm0.004$	& $0.429\pm0.004$		\\
$N_{dbb}$           &$1668^{+112}_{-105}$   &$1990^{+130}_{-120}$&$2628^{+170}_{-150}$\\
$\Gamma$      & $2.85^{+0.33}_{-0.32}$     & 2.85		&	2.85	\\
$N_{po}^b$             & $0.54^{+0.43}_{-0.24}$ &$0.61\pm0.03$	&$0.64\pm0.03$\\ \hline
\end{tabular}
$^a$Column density in units of $10^{22}$\,cm$^{-2}$. \\
$^b$Power-law norm in units of $10^{-1}$ph cm$^{-2}$s$^{-1}$keV$^{-1}$, following \citet{Soria11}
\end{table}

We then include the dust scattering term using a MRN77 dust model with the extraction radius fixed at 45 arcsec and the column density fixed at $7.5\times10^{21}$\,cm$^{-2}$, from the LAB survey \citep{LAB05}.  The only potentially free parameter is the plane of the dust.  At an assumed distance of 8 kpc and Galactic latitude of $-1.36^{\circ}$, GRS1759-258 is $\sim 190$\,pc out of the plane, while the scale height for cold molecular clouds is $\sim 100$\,pc \citep{Cox05}.  Fitting with a variable dust position places the dust at $x>0.91$\ of the distance to the source. At this position, the effect of dust scattering is minimal and the best-fit parameters are essentially identical to the model without dust scattering.  While this is a possible scenario, we considered more likely a case where the dust cloud position is fixed halfway between the source and observer ($x=0.5$) -- putting the cloud at approximately the scale height of cold clouds in the ISM.  These are the values shown in Table~\ref{tab:GRS1758}.  Due to the small-angle nature of dust scattering, any value between $x=0$\ and 0.5 would give essentially the same result, as shown in Figure~\ref{fig:sigma}[Right].  With this dust position, we see that the best-fit intrinsic column density is significantly lower, the disk blackbody temperature drops slightly, and the best-fit disk normalization increased significantly. 

As with XTEJ1701-462, these changes affect the inferred parameters of the system at the 1-2$\sigma$\ level.  \citet{Soria11} noted that a 2003 INTEGRAL study found $N_{dbb} \approx 2700$\ during another soft state period, which compared poorly to the $1668$\ value they found, leading them to use an average value of 2200 in their calculations.  However, when using the latest calibration and including a reasonable dust scattering model, we find a disk normalization in good agreement with the INTEGRAL results, leading to a $\sqrt{2700/2100} \sim 10$\% increase in the inferred BH mass. 

\subsection{Burst sources: EXO 0748-676 redux}

We have already described fits to EXO 0748-676 above in the context of measuring the cooling of the neutron star after a long outburst.  However, this source is extremely complex, a transient low mass X-ray binary system that exhibits bursts, dips, and eclipses in its light curve. 
Numerous studies \citep[\protect{{\it e.g.}}][]{Church98, BB01, Homan03} have been conducted that fit the spectrum during dipping and non-dipping states, but the interpretation of these fits is a source of debate. More recently, \citet{AD06} jointly fit XMM spectra during burst and quiescent states and found that a combination of a partially covered power law, bremsstrahlung, and blackbody (for the burst emission) produced good fits.

We re-analyzed the first data set listed by \citet{AD06} in their Table 1, that is, XMM-Newton ObsID 0123500101. We reprocessed the data using SAS v14.0, then extracted and fitted the spectra for the persistent emission and a burst (Burst VII, as designated by \citet{Homan03}) simultaneously using the procedures described in \citet{AD06}. Table \ref{tab:EXO0748} lists our results, with those of \citet{AD06} for comparison. We then included the dust scattering term and refit the spectra. As with the GRS1758-258 fit, only the plane of the dust was allowed to float. This produced a fit that placed the cloud at about 90\% of the distance to the source. However, the source is about 8$\pm$1 kpc  away \citep{Jonker04}, and with a Galactic latitude of -19.81$^{\circ}$, this means that the cloud has the unlikely height of $|z| \sim 2.4$ kpc above the Galactic plane. If the cloud altitude is held at the more realistic value of 100 pc, the fits are very similar to those found without the scattering term; these are shown in Table \ref{tab:EXO0748}. The main difference is in the parameters associated with the in-system absorption during the persistent state, with the partial covering fraction dropping from 0.95$\pm$0.02 to 0.89$\pm$0.01, and the hydrogen column density falling from $8\pm1 \times 10^{22}$\,cm$^{-2}$ to $7\pm1\times10^{22}$\,cm$^{-2}$.

\begin{table}
\begin{center}
\caption{Best-fit spectral parameters for EXO 0748-676 in the Persistent and Burst States \label{tab:EXO0748}}
\begin{tabular}{l lll}
\tableline \tableline
Parameter           & \multicolumn{3}{c}{Persistent, Burst}     \\
                    & AD(06)$^{a}$           & This paper  & +xscat         \\
\tableline
$N_H$(gal, abs)$^{b,c}$ & 0.11           & 0.11            & 0.11                \\
$N_H$(gal, scat)$^{b}$    &  ...             & ...             & 0.11                 \\
$kT_{bremss}$       & 0.6$\pm$0.1      & 0.3$\pm$0.1     & 0.3$\pm$0.1    \\
$\Gamma$            & 1.2$\pm$0.1      & 1.2$\pm$0.1     & 1.1$\pm$0.1    \\
$kT_{bb}$           & ..., 1.7$\pm$0.1 & ..., 1.8$\pm$0.1& ..., 1.8$\pm$0.1   \\
$N_H$(part. cov.)$^{b}$   & 8$\pm$1,2$\pm$1  & 8$\pm$1,2$\pm$1 & 7$\pm$1,2$\pm$2         \\
PC fraction         & 0.98$\pm$0.01,   &0.95$\pm$0.02,   & 0.89$\pm$0.01, \\
                    & 0.4$\pm$0.1      &0.5$\pm$0.2      & 0.4$\pm$0.2 \\ \hline
\end{tabular}
%\tablenotetext{a}\citet{AD06}.}
\end{center}
$^a$\,\protect{\citet{AD06}} \\
$^b$\,All column densities in units of $10^{22}$\,cm$^{-2}$ \\
$^{c}$\,{Galactic absorption was held at the value found by Dickey \& Lockman (1990).}
\end{table}

\section{Conclusions}

We have described  a new dust scattering model and code, \textit{xscat}, that includes not only accurate Mie scattering, but also a wide range of current dust models with the capability of easily adding new models as they are developed.  This code calculates the scattering of X-rays as a function of angle for a single cloud, including the effects of realistic extraction circles which would include some scattered photons.  The XSPEC \textit{xscat} model uses output from the \textit{xscat} code to determine the X-ray scattering cross section due to dust as a function of energy, using as input parameters the desired dust model and the extraction region used for the source.  The two variable parameters are the hydrogen column density N$_{\rm H}$ (a proxy for the total dust column density, for a given dust model) and the relative position of the dust along the line of sight. Including an absorption model such as `phabs', which based on the \citet{Verner96} data or `tbabs', which uses updated higher-resolution cross sections \citep{Wilms00} together with this dust scattering model and tying together the values of N$_{\rm H}$, ensures a consistent measurement of extinction in spectral fits.

Applying these calculations to existing analyses shows that while the effects are modest, they are both significant and systematic and impact even lightly-absorbed low-resolution spectra. These results also show the impact of the choice of extraction region both for the source and background. When dust scattering is significant, using an annulus around the source as a background region may be inappropriate since this is precisely the region where source photons will be scattered.  The \textit{xscat} model can be used to determine the impact of this effect by exploring the change in scattering cross section for a range of  extraction regions. 

The cross section for dust scattering is smaller than absorption for $E<6$\,keV, but has a different energy dependence.  As the photon energy increases, atomic absorption cross sections scale as $\propto E^{-3}$\ to $E^{-3.5}$, while dust scattering is $\propto E^{-2}$, making the effect of dust scattering on spectral fits hard to predict {\it a priori}.  As the scattering cross section is between 5-20\% of absorption for energies $< 2$\,keV, the effects will usually be modest except in  situations with extremely large column densities.  

The effect of dust scattering may also be significant for detectors with limited spatial resolution, if the underlying X-ray source is time-variable.  Depending upon the exact position of dust grains along the line of sight and the pathlength differences created by the small scattering angles, a source that suddenly brightens by an order of magnitude for example will see a similar level of enhancement in the halo on scales of days to weeks later \citep[\protect{{\it e.g.}}][]{Vaughan04}.  Thus even non-imaging detectors such as RXTE PCA or the Astrosat LAXPC could be affected when comparing observations of a flaring or dipping source. 

Finally, we note that dust scattering also shows distinct features at atomic edges which could be resolved in detectors with $\Delta E < 10$\,eV.  These calculations, however, are only for dust scattering; solid-state X-ray Absorption Fine Structure (XAFS) \citep{Lee09} features, for example, should also be included when considering high-resolution spectra.  

\acknowledgements

The authors thank the anonymous referee for prompt and helpful comments that significantly improved the work. We also thank Sebastian Heinz for reviewing the xscat code and helping to debug it. Randall Smith gratefully acknowledges helpful discussions and overall inspiration to work on X-ray scattering from Eli Dwek.  Financial support for this work was made possible by Chandra grant TM4-15002X. %{\it Facilities:} \facility{CXO}, \facility{XMM-Newton}.

\end{document}